\documentclass{article}

\usepackage{jason}
\usepackage{color}
\usepackage{paralist}
\usepackage{version}
\usepackage{url}
\usepackage{tikz}
\usetikzlibrary{positioning,arrows,automata,shapes,decorations.pathreplacing}

\usepackage[a4paper]{geometry}

\newcommand{\wsp}{{\sc WSP}}
\newcommand{\wspi}[1]{\text{WSP}_{#1}}
\newcommand{\owsp}{{\sc OWSP}}
\newcommand{\rrbac}{{\sf R$^2$BAC}}

\newenvironment{oldstuff}{\color{red}}{}

\excludeversion{oldstuff}

\begin{document}

\title{Constraint Expressions and Workflow Satisfiability}
\author{Jason Crampton \and Gregory Gutin}
\maketitle

\begin{abstract}
A workflow specification defines a set of steps and the order in which those steps must be executed.
Security requirements and business rules may impose constraints on which users are permitted to perform those steps.
A workflow specification is said to be satisfiable if there exists an assignment of authorized users to workflow steps that satisfies all the constraints.
An algorithm for determining whether such an assignment exists is important, both as a static analysis tool for workflow specifications, and for the construction of run-time reference monitors for workflow management systems.
We develop new methods for determining workflow satisfiability based on the concept of constraint expressions, which were introduced recently by Khan and Fong.
These methods are surprising versatile, enabling us to develop algorithms for, and determine the complexity of, a number of different problems related to workflow satisfiability.
\end{abstract}

\section{Introduction}\label{sec:intro}

It is increasingly common for organizations to computerize their business and management processes.
The co-ordination of the tasks or steps that comprise a computerized business process is managed by a workflow management system (or business process management system).
Typically, the execution of these steps will be triggered by a human user, or a software agent acting under the control of a human user, and the execution of each step will be restricted to some set of authorized users.

A workflow is defined by the steps that comprise a business process and the order in which those steps should be performed.
Moreover, it is often the case that some form of access control, often role-based, should be applied to limit the execution of steps to authorized users.
In addition, many workflows require controls on the users that perform groups of steps.
The concept of a Chinese wall, for example, limits the set of steps that any one user can perform~\cite{BrNa89}, as does separation-of-duty, which is a central part of the role-based access control model~\cite{ansi-rbac04}.
Hence, it is important that workflow management systems implement security controls that enforce authorization rules and business rules, in order to comply with statutory requirements or best practice~\cite{BaBuKa10}.
It is these ``security-aware'' workflows that will be the focus of the remainder of this paper.

\begin{figure}[h]\centering
\subfigure[Ordering on steps]{
\begin{tikzpicture}[->,.=stealth',node distance=8mm and 7mm,semithick,auto]
  \node  (t1)                      {$s_1$};
  \node  (t2) [right=of t1]        {$s_2$};
  \node  (t3) [above right=of t2]  {$s_3$};
  \node  (t4) [below right=of t3]  {$s_4$};
  \node  (t5) [above right=of t4]  {$s_5$};
  \node  (t6) [below right=of t5]  {$s_6$};
  \path (t1) edge (t2)
   (t2) edge (t3)
   (t2) edge (t4)
   (t3) edge (t5)
   (t4) edge (t6)
   (t5) edge (t6);
\end{tikzpicture}}
\hfill
\subfigure[Constraints]{
\begin{tikzpicture}[-,node distance=8mm and 8mm,semithick,auto]
  \node (t1) {$s_1$};
  \node (t2) [above=of t1] {$s_2$};
  \node (t3) [left=of t1] {$s_3$};
  \node (t4) [right=of t1] {$s_4$};
  \node (t5) [left=of t3] {$s_5$};
  \node (t6) [right=of t4] {$s_6$};
  \path (t1) edge [dotted] node {$=$} (t3)
        (t3) edge [dotted] node {$\ne$} (t5)
        (t1) edge [dotted] node[swap] {$\ne$} (t4)
        (t1) edge [dotted] node[swap] {$\ne$} (t2)
        (t4) edge [dotted] node[swap] {$\ne$} (t6);
\end{tikzpicture}}
\hfill
\subfigure[Legend]{\footnotesize\setlength{\extrarowheight}{2pt}
  \begin{tabular}{|ll|}
    \hline
    $s_1$ & create purchase order \\
    $s_2$ & approve purchase order \\
    $s_3$ & sign GRN \\
    $s_4$ & create payment \\
    $s_5$ & countersign GRN \\
    $s_6$ & approve payment \\
    \hline
    $\ne$ & different users must perform steps \\
    $=$ & same user must perform steps \\
    \hline
  \end{tabular}}
\caption{A simple constrained workflow for purchase order processing}\label{fig:example-workflow}
\end{figure}
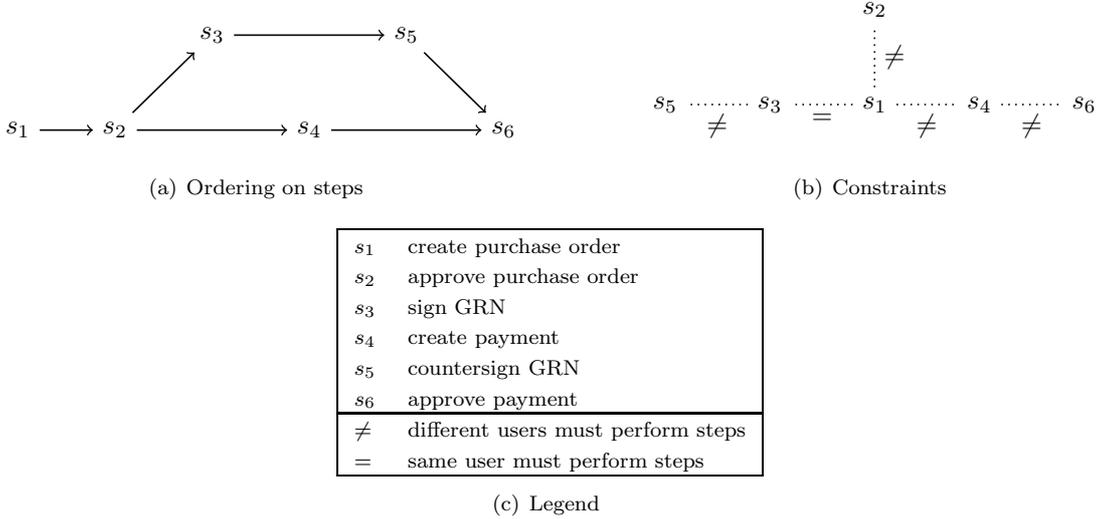

A simple, illustrative example for purchase order processing~\cite{cram:sacmat05} is shown in Figure~\ref{fig:example-workflow}.
In the first step of the workflow, the purchase order is created and approved (and then dispatched to the supplier).
The supplier will submit an invoice for the goods ordered, which is processed by the create payment step.
When the supplier delivers the goods, a goods received note (GRN) must be signed and countersigned.
Only then may the payment be approved and sent to the supplier.
Note that a workflow specification need not be linear: the processing of the GRN and of the invoice can occur in parallel, for example.

In addition to defining the order in which steps must be performed, the workflow specification includes rules to prevent fraudulent use of the purchase order processing system.
In our example, these rules restrict the users that can perform pairs of steps in the workflow: the same user may not sign and countersign the GRN, for example.

It is apparent that it may be impossible to find an assignment of authorized users to workflow steps such that all constraints are satisfied.
In this case, we say that the workflow specification is \emph{unsatisfiable}.
The {\sc Workflow Satisfiability Problem} (\wsp) is known to be NP-hard, even when the set of constraints only includes constraints that have a relatively simple structure (and that would arise regularly in practice).%
\footnote{In particular, the {\sc Graph $k$-Colorability} problem can be reduced to a special case of \wsp\ in which the workflow specification only includes separation-of-duty constraints~\cite{WangLi10}.}

The rules described above can be encoded using constraints~\cite{cram:sacmat05}, the rules being enforced if and only if the constraints are satisfied.
More complex constraints, in which restrictions are placed on the users who execute sets of steps can also be defined~\cite{ArPo09,CrGuYe12,WangLi10}, can encode more complex business requirements.
(We describe these constraints in more detail in Section~\ref{sec:wsp}.)
A considerable body of work now exists on the satisfiability of workflow specifications that include such constraints~\cite{BaBuKa10,CrGuYe12,WangLi10}.

In this paper, we use constraint expressions to solve \wsp.
Constraint expressions were introduced by Khan and Fong in their work on workflow feasibility~\cite{KhFo12}.
However, the potential of constraint expressions was not fully realized.
In this paper, we show how constraint expressions can be used to solve \wsp\ and a number of related problems.

We also introduce a set of operators for combining workflows.
This allows us to model workflows in which the execution of steps is determined at execution time, which we will call \emph{conditional workflows}.
Our model enables us to formulate the satisfiability problem for conditional workflows, which we solve using constraint expressions.
To our knowledge, these are the first results on conditional workflows.

The main contributions of this paper are:
  \begin{itemize}
    \item to generalize the results of Wang and Li on the fixed parameter tractability of \wsp\ (Section~\ref{sec:constraint-expressions});
    \item to introduce a language for workflow composition (Section~\ref{sec:conditional-wsp});
    \item to establish new results on the satisfiability of conditional workflows (Section~\ref{sec:conditional-wsp});
    \item to demonstrate how a problem studied by Armando {\em et al.}~\cite{ArPo09} and a problem introduced by Crampton~\cite{cram:sacmat05} can be solved using constraint expressions (Section~\ref{sec:other-applications}).
  \end{itemize}

In the next section we provide relevant background material.
In Section~\ref{sec:constraint-expressions}--\ref{sec:other-applications}, we describe our results.
The proofs of our results can be found in the appendix, with the exception of the proof of Theorem~\ref{thm:wsp2-elementary}.
We conclude with a summary of our contributions, a discussion of related work, and our plans for future work.

\section{Background}\label{sec:background}

In this section, we introduce our notation and definitions, derived from earlier work~\cite{cram:sacmat05,WangLi10}, and then define the workflow satisfiability problem.
In order to make the paper self-contained, we also provide a short overview of parameterized complexity and summarize a number of useful results from the literature.

\subsection{The Workflow Satisfiability Problem}\label{sec:wsp}

A directed acyclic graph $G = (V,E)$ is defined by a set of nodes $V$ and a set of edges $E \subseteq V \times V$.
The reflexive, transitive closure of a directed acyclic graph defines a partial order, where $v \leqslant w$ if and only if there is a path from $v$ to $w$ in $G$.
If $(V,\leqslant)$ is a partially ordered set, then we write $v \parallel w$ if $v$ and $w$ are incomparable; that is, $v \not\leqslant w$ and $w \not\leqslant v$.
We may write $v \geqslant w$ whenever $w \leqslant v$.
We may also write $v < w$ whenever $v \leqslant w$ and $v \ne w$.
Finally, we will write $[n]$ to denote $\set{1,\dots,n}$.

\begin{Def}\label{def:workflow}
A \emph{workflow specification} is defined by a directed, acyclic graph $G = (S,E)$, where $S$ is a set of steps and $E \subseteq S \times S$.
Given a workflow specification $(S,E)$ and a set of users $U$, an \emph{authorization policy} for a workflow specification is a relation $A \subseteq S \times U$.
A \emph{workflow authorization schema} is a tuple $(G,U,A)$, where $G = (S,E)$ is a workflow specification and $A$ is an authorization policy.
\end{Def}

We will use the representations of a workflow specification as a partial order and a DAG interchangeably.
The workflow specification describes a sequence of steps and the order in which they must be performed when the workflow is executed, each such execution being called a \emph{workflow instance}.
If $s < s'$ then $s$ must be performed before $s'$ in every instance of the workflow; if $s \parallel s'$ then $s$ and $s'$ may be performed in either order.
User $u$ is authorized to perform step $s$ only if $(s,u) \in A$.%
\footnote{In practice, the set of authorized step-user pairs, $A$, will not be defined explicitly.
          Instead, $A$ will be inferred from other access control data structures.
          In particular, \rrbac~--~the role-and-relation-based access control model of Wang and Li~\cite{WangLi10}~--~introduces a set of roles $R$, a user-role relation ${\it UR} \subseteq U \times R$ and a role-step relation ${\it SA} \subseteq R \times S$ from which it is possible to derive the steps for which users are authorized.
          For all common access control policies (including \rrbac), it is straightforward to derive $A$.
          We prefer to use $A$ in order to simplify the exposition.}
We assume that for every step $s \in S$ there exists some user $u \in U$ such that $(s,u) \in A$.

\begin{Def}
Let $((S,E),U,A)$ be a workflow authorization schema.
A \emph{plan} is a function $\pi: S \rightarrow U$.
A plan $\pi$ is \emph{authorized} for $((S,E),U,A)$ if $(s,\pi(s)) \in A$ for all $s \in S$.
\end{Def}

\begin{Def}
A \emph{workflow constraint} has the form $(\rho,S_1,S_2)$, where $S_1,S_2 \subseteq S$ and $\rho \subseteq U \times U$.
A \emph{constrained workflow authorization schema} is a tuple $((S,E),U,A,C)$, where $C$ is a set of workflow constraints.
\end{Def}

\begin{Def}
A plan $\pi : S \rightarrow U$ \emph{satisfies} a workflow constraint $(\rho,S_1,S_2)$ if there exist $s_1 \in S_1$ and $s_2 \in S_2$ such that $(\pi(s_1),\pi(s_2)) \in \rho$.
Given a constrained workflow authorization schema $((S,E),U,A,C)$, a plan $\pi$ is \emph{valid} if it is authorized and it satisfies all constraints in $C$.
\end{Def}

We write $\Delta \subseteq U \times U$ to denote the diagonal relation $\set{(u,u) : u \in U}$ and $\Delta^c$ to denote its complement $\set{(u,u) : (u,u) \not\in \Delta}$.
Thus, the constraint on steps $s_1$ and $s_2$ in Figure~\ref{fig:example-workflow} would be written as $(\Delta^c,\set{s_1},\set{s_2})$.
%

We may now define the workflow satisfiability problem, as defined by Wang and Li~\cite{WangLi10}.

\begin{center}
\fbox{%
      \begin{tabulary}{.95\columnwidth}{@{}r<{~}@{}L@{}}
        \multicolumn{2}{@{}l}{\sc Workflow Satisfiability Problem (WSP)}\\
        \emph{Input:} & A constrained workflow authorization schema $((S,E),U,A,C)$\\
        \emph{Output:} & A valid plan $\pi : S \rightarrow U$ or an answer that there exists no valid plan
       \end{tabulary}%
      }
\end{center}

We now discuss constraints in more detail, including the type of business rules we can encode using our constraints and compare them to constraints in the literature.
Our definition of workflow constraint is more general than similar definitions used when studying \wsp.
Crampton defined constraints in which $S_1$ and $S_2$ are singleton sets: we will refer to constraints of this form as \emph{Type $1$} constraints; for brevity we will write $(\rho,s_1,s_2)$ for the Type 1 constraint $(\rho,\set{s_1},\set{s_2})$.
Wang and Li defined constraints in which at least one of $S_1$ and $S_2$ is a singleton set: we will refer to constraints of this form as \emph{Type $2$} constraints and we will write $(\rho,s_1,S_2)$ in preference to $(\rho,\set{s_1},S_2)$.
Constraints in which $S_1$ and $S_2$ are arbitrary sets will be called \emph{Type $3$} constraints.

We say that two constraints $\gamma$ and $\gamma'$ are \emph{equivalent} if a plan $\pi$ satisfies $\gamma$ if and only if it satisfies $\gamma'$.
The Type 2 constraint $(\rho,s_1,S_2)$ is equivalent to $(\rho,S_2,s_1)$ if $\rho$ is symmetric, in which case we will write $(\rho,s_1,S_2)$ in preference to $(\rho,S_2,s_1)$.

It is worth pointing out that Type 1 constraints can express requirements of the form described in Section~\ref{sec:intro}, where we wish to restrict the combinations of users that perform pairs of steps.
The plan $\pi$ satisfies constraint $(\Delta,s,s')$, for example, if the same user is assigned to both steps by $\pi$, and satisfies constraint $(\Delta^c,s,s')$ if different users are assigned to $s$ and $s'$.
In other words, these represent, respectively, binding-of-duty and separation-of-duty constraints.
Abusing notation in the interests of readability, we will replace $\Delta$ and $\Delta^c$ by $=$ and $\ne$, respectively.

Type 2 constraints provide greater flexibility, although Wang and Li, who introduced these constraints, do not provide a use case for which such a constraint would be needed.
However, there are forms of separation-of-duty requirements that are most naturally encoded using Type 3 constraints.
Consider, for example, the requirement that a set of steps $S' \subseteq S$ must not all be performed by the same user~\cite{ArGiPo09}.
We may encode this as the constraint $(\ne,S',S')$, which is satisfied by a plan $\pi$ only if there exists two steps in $S'$ that are allocated to different users by $\pi$.

Henceforth, we will write \wsp$(\rho_1,\dots,\rho_t)$ to denote a special case of \wsp\ in which all constraints have the form $(\rho_i,S',S'')$ for some $\rho_i \in \set{\rho_1,\dots,\rho_t}$ and for some $S',S'' \subseteq S$.
We will write $\wspi{i}(\rho_1,\dots,\rho_t)$ to denote a special case of \wsp$(\rho_1,\dots,\rho_t)$, in which there are no constraints of Type $j$ for $j > i$.
Thus, $\wspi{1}(=,\ne)$, for example, indicates an instance of \wsp\ in which all constraints  have the form \mbox{$(=,s_1,s_2)$} or $(\ne,s_1,s_2)$ for some $s_1,s_2 \in S$.

We will write $c$, $n$ and $k$ to denote the number of constraints, users and steps, respectively, in an instance of \wsp.
We will analyze the complexity of the workflow satisfiability problem in terms of these parameters.

Note that definition of \wsp\ given above does not make any reference to the ordering on the set of steps.
The original definition, as formulated by Crampton~\cite{cram:sacmat05}, included constraints that were sensitive to the order in which steps were executed.
If $s \parallel s'$, we may define two different constraints $(\rho,s,s')$ and $(\rho',s',s)$, the first of which must be satisfied if $s$ is performed before $s'$, while the second must be satisfied if $s'$ is performed before $s$.
To facilitate direct comparison with the work of Wang and Li on \wsp, we defer the analysis of Crampton's version of the problem until Section~\ref{sec:other-applications}.

\subsection{Applications of \wsp}\label{sec:applications-of-wsp}

There are a number of different execution models for workflow systems.
In some systems, a tasklist is created when a workflow is instantiated.
The tasklist is simply a valid plan for the worfklow instance, allocating users to specific steps in the workflow instance.
In other systems, the workflow system maintains a pool of ready steps for each worfklow instance.
We say a step is \emph{ready} in a workflow instance if all its immediate predecessor steps have been executed.
The workflow system may allocate ready steps to users; alternatively users may select steps to perform from the pool.
In both cases, the system must ensure both that the user is authorized and that allowing the user to perform the step does not prevent the remaining steps in the workflow instance from completing.

For systems that create tasklists, it is sufficient to know that the workflow specification is satisfiable.
Thus, an algorithm for deciding \wsp\ is an important static analysis tool for such systems.
However, such an algorithm will only need to be executed when the workflow specification is created or when it changes.
The fact that the problem is NP-hard means that it is important to find as efficient an algorithm as possible.

For other systems, however, the algorithm will need to be run repeatedly: every time a user is allocated to a step.
Note that the decision whether to allow a user to execute a step in a partially completed workflow instance can be determined by solving an instance of \wsp.
Specifically, suppose $W = ((S,E),U,A,C)$ is a workflow specification, some subset $S'$ of steps have been performed in some instance of $W$, and the system needs to decide whether to allow $u'$ to perform $s'$.
Thus we have a partial plan $\pi : S' \rightarrow U$.
We then construct a new workflow instance $W' = ((S,E),U,A',C)$, where $(s,u) \in A'$ if and only if one of the following conditions holds:
  \begin{inparaenum}[(i)]
    \item $s \in S'$ and $u = \pi(s)$
    \item $s = s'$ and $u = u'$
    \item $s \not\in S' \cup \set{s'}$ and $(u,s) \in A$.
  \end{inparaenum}
Clearly, the workflow instance is satisfiable (when $u'$ performs $s'$) if and only if $W'$ is satisfiable.
Assuming that these checks should incur as little delay as possible, particularly in the case when users select steps in real time~\cite{KoSc08}, it becomes even more important to find an algorithm that can decide \wsp\ as efficiently as possible.

The definition of workflow satisfiability given above assumes that the set of users and the authorization relation are given.
This notion of satisfiability is appropriate when the workflow schema is designed ``in-house''.
A number of large information technology companies develop business process systems which are then configured by the end users of those systems.
Part of that configuration includes the assignment of users to steps in workflow schemas.
The developer of such a schema may wish to be assured that the schema is satisfiable for some set of users and some authorization relation, since the schema is of no practical use if no such user set and authorization relation exist.
The desired assurance can be provided by solving an instance of \wsp\ in which there are $k$ users, each of which is authorized for all steps.
The developer may also determine the minimum number of users required for a workflow schema to be satisfiable.
The minimum number must be between $1$ and $k$ and, using a binary search, can be determined by examining $\ceil{\log_2 k}$ instances of \wsp.

\vfill

\subsection{Parameterized Complexity}\label{sec:fpt}

A na\"ive approach to solving \wsp\ would consider every possible assignment of users to steps in the workflow.
There are $n^k$ such assignments if there are $n$ users and $k$ steps, so an algorithm of this form would have complexity $O(cn^k)$, where $c$ is the number of constraints.
Moreover, Wang and Li showed that \wsp\ is NP-hard, by reducing {\sc Graph $k$-Colorability} to \wsp$(\ne)$~\cite[Lemma 3]{WangLi10}.
In short, \wsp\ is hard to solve in general.
The importance of finding an efficient algorithm for solving \wsp\ led Wang and Li to look at the problem from the perspective of parameterized complexity~\cite{WangLi10}.

Suppose we have an algorithm that solves an NP-hard problem in time $O(f(k)n^d)$, where $n$ denotes the size of the input to the problem, $k$ is some (small) parameter of the problem, $f$ is some function in $k$ only, and $d$ is some constant (independent of $k$ and $n$).
Then we say the algorithm is a \emph{fixed-parameter tractable} (FPT) algorithm.
If a problem can be solved using an FPT algorithm then we say that it is an \emph{FPT problem} and that it belongs to the class FPT~\cite{DowneyFellows99,Ni06}.

Wang and Li showed, using an elementary argument, that $\wspi{2}(\ne)$ is FPT and can be solved in time $O(k^{k+1}N)$, where $N$ is the size of the entire input to the problem~\cite[Lemma 8]{WangLi10}.
They also showed that \mbox{$\wspi{2}(\ne,=)$} is FPT~\cite[Theorem 9]{WangLi10}, using a rather more complex approach: specifically, they constructed an algorithm that runs in time \mbox{$O(k^{k+1}(k-1)^{k2^{k-1}}N)$}; it follows that $\wspi{2}(=,\ne)$ is FPT.
One of the contributions of this paper is to describe a new method for solving $\wspi{3}(=,\ne)$ (that can also be used to solve $\wspi{2}(=,\ne)$), thus generalizing Wang and Li's result.

\begin{oldstuff}
When the runtime $O(f(k)n^d)$ is replaced by the much more powerful $O(n^{f(k)})$, we obtain the class XP, where each problem is polynomial-time solvable for any fixed value of $k$.
There is an infinite collection of parameterized complexity classes, $\text{W}[1], \text{W}[2],\dots$, with $\text{FPT} \subseteq \text{W[1]} \subseteq \text{W[2]} \subseteq \dots \subseteq \text{XP}$.

Informally, a parameterized problem belongs to the complexity class W[$i$] if there exists an FPT algorithm that transforms every instance of the problem into an instance of {\sc Weighted Circuit Satisfiability} for a circuit of weft $i$.
It can be shown that FPT is the class W[$0$].
The problems {\sc Independent Set} and {\sc Dominating Set} are in W[1] and W[2], respectively.
It is widely-believed and often assumed that $\text{FPT} \neq \text{W}[1]$.
For a more formal introduction to the W family of complexity classes, see Flum and Grohe~\cite{FlumGrohe06}.

Wang and Li~\cite[Theorem 10]{WangLi10} proved that WSP (for arbitrary relations defined on the user set) is W[1]-hard in general, using a reduction from {\sc Independent Set}.
By definition, FPT is a subset of W[1] and a parameterized analog of Cook's Theorem~\cite{DowneyFellows99} as well as the Exponential Time Hypothesis~\cite{FlumGrohe06,ImPaZa01} strongly support the widely held view that FPT is not equal to W[1].
One of the main contributions of this paper is to extend the set of special cases of WSP that are known to be FPT.
\end{oldstuff}

\section{Solving WSP Using Constraint Expressions}\label{sec:constraint-expressions}

In this section, we show how to extend the elementary methods used by Wang and Li to obtain results for \mbox{$\wspi{2}(=,\ne)$} and $\wspi{3}(=,\ne)$.
Informally, our results make use of two observations:
  \begin{itemize}
    \item A construction used by Crampton {\em et al.}~\cite{CrGuYe12a} can be used to transform an instance of $\wspi{1}(=,\ne)$ into an equivalent instance of \mbox{$\wspi{1}(\ne)$} in time polynomial in the numbers of constraints, steps and users.
    \item We can transform an instance of $\wspi{i}(=,\ne)$ into multiple instances of $\wspi{1}(=,\ne)$, the number of instances being dependent only on the number of steps.
  \end{itemize}
We use constraint expressions~\cite{KhFo12} to represent workflow constraints and to reason about multiple constraints and the relationships between different types of constraints.

\subsection{Reducing $\wspi{1}(=,\ne)$ to $\wspi{1}(\ne)$}

The basic idea is to merge all steps that are related by constraints of the form $(=,s_1,s_2)$ for $s_1,s_2 \in S$.
More formally, consider an instance $\cal I$ of $\wspi{1}(=,\ne)$, given by a workflow $((S,E),U,A,C)$.
\begin{enumerate}[(1)]
  \item Construct a graph $H$ with vertices $S$, in which $s',s''\in S$ are adjacent if $C$ includes a constraint $(=,s',s'')$.
  \item If there is a connected component of $H$ that contains both $s'$ and $s''$ and $C$ contains a constraint $(\ne,s',s'')$ then $\cal I$ is unsatisfiable, so we may assume there is no such connected component.
  \item For each connected component $T$ of $H$,
    \begin{enumerate}[(a)]
      \item replace all steps of $T$ in $S$ by a ``superstep'' $t$;
      \item for each superstep $t$, authorize user $u$ for $t$ if and only if $u$ was authorized (by $A$) for all steps in $t$
      \item for each such superstep $t$, merge all constraints for steps in $t$.
    \end{enumerate}
\end{enumerate}
Clearly, we now have an instance of $\wspi{1}(\ne)$, perhaps with fewer steps and a modified authorization relation, that is satisfiable if and only if $\cal I$ is satisfiable.
For ease of reference, we will refer to the procedure described above as the \emph{$\wspi{1}$ constraint reduction method}.
The reduction can be performed in time $O(kc + kn)$, where $c$ is the number of constraints: step (1) takes time $O(k+c)$; step (3) performs at most $k$ merges; each merge takes $O(k+c+n)$ time (since we need to merge vertices, and update constraints and the authorization relation for the new vertex set);\footnote{We can check step (2) when we merge constraints in step 3(c).} finally, if $k \leqslant c$ we have $O(k(k+c+n) = O(k(c+n))$, and if $c \leqslant k$ then we perform no more than $c$ merges in time $O(c(k+c+n)) = O(ck+cn) = O(ck+kn)$.

\subsection{Constraint Expressions}

To understand the intuition behind our approach, consider a workflow \mbox{$W = ((S,E),U,A,\set{(\rho,S',S'')})$}, which defines an instance of $\wspi{3}(\rho)$.
By definition, a plan $\pi$ satisfies the constraint $(\rho,S',S'')$ if there exist $s' \in S'$ and $s'' \in S''$ such that $(\pi(s'),\pi(s'')) \in \rho$.
In other words, we could decide the satisfiability of $W$ by considering the satisfiability of multiple instances of $\wspi{1}$: specifically, for each pair $(s',s'') \in S' \times S''$, we consider the satisfiability of the workflow $((S,E),U,A,\set{(\rho,s',s'')})$; if any one of these instances is satisfiable, then so is $W$.
On the other hand, a plan satisfies a workflow $W = ((S,E),U,A,\set{\gamma_1,\gamma_2})$, for constraints $\gamma_1$ and $\gamma_2$, if and only $\pi$ satisfies workflows $((S,E),U,A,\set{\gamma_1})$ and $((S,E),U,A,\set{\gamma_2})$.

More formally, given a set of steps $S$, we define a \emph{constraint expression} recursively:
  \begin{itemize}
    \item $(\rho,s_1,s_2)$ is a (\emph{primitive}) constraint expression;
    \item if $\gamma$ and $\gamma'$ are constraint expressions, then $\gamma \wedge \gamma'$ and $\gamma \vee \gamma'$ are constraint expressions.
  \end{itemize}
A plan $\pi$ satisfies constraint expression:
\begin{itemize}
  \item $\gamma \wedge \gamma'$ if and only if $\pi$ satisfies $\gamma$ and $\gamma'$; and
  \item $\gamma \vee \gamma'$ if and only if $\pi$ satisfies $\gamma$ or $\gamma'$.
\end{itemize}

\subsection{Reducing $\wspi{}(\rho_1,\dots\rho_t)$ to $\wspi{1}(\rho_1,\dots,\rho_t)$}

We now express workflow specifications using constraint expressions, rather than sets of constraints.
A constraint $(\rho,S',S'')$, $\rho \in \set{\rho_1,\dots,\rho_t}$, is equivalent to a constraint expression $\bigvee_{s' \in S',s'' \in S''}(\rho,s',s'')$, so every constraint can be written as the disjunction of primitive constraints.
Moreover, the set of constraints $\set{\Gamma_1,\dots,\Gamma_c}$, where each $\Gamma_i$ is a disjunction of primitive constraint expressions, is equivalent to the constraint expression $\Gamma_1 \wedge \dots \wedge \Gamma_c$.

In other words, we can reduce the problem of determining the satisfiability of \mbox{$((S,E),U,A,C)$} to the problem of determining the satisfiability of a workflow of the form
  \[
    ((S,E),U,A,\Gamma_1 \wedge \dots \wedge \Gamma_c),
  \]
where $c = \card{C}$; each \emph{clause} $\Gamma_i = (\rho,S'_i,S''_i)$ has the form $\gamma_{i,1} \vee \dots \vee \gamma_{i,m(i)}$, with \mbox{$m(i) = \card{S'_i} \cdot \card{S''_i}$}; and each \emph{literal} $\gamma_{i,j}$ has the form $(\rho,s',s'')$ for some $s' \in S'_i$ and $s'' \in S''_i$.
In other words, we can represent any instance of $\wspi{3}(=,\ne)$ as a workflow containing a constraint expression in ``conjunctive normal form'' in which each of the ``literals'' is a primitive constraint (which corresponds to a single Type 1 constraint).
Moreover, each literal is positive.

\subsection{Solving $\wspi{}(=,\ne)$}\label{sec:solving-wsp3}

Given a constraint expression $\Gamma_1 \wedge \dots \wedge \Gamma_c$, it is easy to see that if we can find a plan $\pi$ for some constraint expression of the form  $\gamma_1 \wedge \dots \wedge \gamma_c$, with $\gamma_i \in \Gamma_i$, then $\pi$ is a plan for $C$.
This is because such a plan satisfies at least one literal in each clause $\Gamma_i$, thereby causing each $\Gamma_i$ to be satisfied; and $C$ is satisfied if each clause is satisfied.
Conversely, if $\pi$ is a plan for $C$ then it is a plan for $\Gamma_1 \wedge \dots \wedge \Gamma_c$ and there exists a workflow expression of the form $\gamma_1 \wedge \dots \wedge \gamma_c$ for which $\pi$ is a plan.
In other words, $\pi$ is a plan for $C$ if and only if it is a plan for $\gamma_1 \wedge \dots \wedge \gamma_c$ for some $\gamma_i \in \Gamma_i$, where $\gamma_i$ is a Type 1 constraint and $\gamma_1 \wedge \dots \wedge \gamma_c$ represents the constraint set $\set{\gamma_1,\dots,\gamma_c}$.
We call $\gamma_1 \wedge \dots \wedge \gamma_c$ a \emph{simple} constraint expression.
That is, we have reduced the satisfiability of an instance of $\wspi{3}(=,\ne)$ to determining the satisfiability of \emph{one or more} instances of $\wspi{1}(=,\ne)$.
The number of instances is equal to $\prod_{i=1}^c \card{\Gamma_i}$, where $\card{\Gamma_i}$ denotes the number of literals (primitive constraint expressions) in $\Gamma_i$.
Our strategy for solving an instance of $\wspi{3}(=,\ne)$, therefore, is to try to determine the satisfiability of these related instances of $\wspi{1}(=,\ne)$.

\begin{Thm}\label{thm:wsp2-elementary}
\wsp$_2(=,\ne)$ and \wsp$_3(=,\ne)$ can be decided in time
\[
  O((k-1)^c(c(k-1)^k + kn))\quad\text{and}\quad O\Big(k^{2c}(c(k-1)^k + kn)\Big),
\]
respectively, where $c$ is the number of constraints in the workflow instance.
\end{Thm}

\begin{proof}
We first consider an instance of $\wspi{1}(=,\ne)$, to which we apply the $\wspi{1}$ constraint reduction to obtain an instance of $\wspi{1}(\ne)$.
As any step with at least $k$ authorized users can be assigned a user that has not been assigned to any other step, we may focus on the allocation of users to steps having fewer than $k$ authorized users.

We consider each possible plan in turn and for each plan we check whether every constraint is satisfied.
There are no more than $(k-1)^k$ plans to check---since each of the steps has at most $k-1$ authorized users and there are no more than $k$ steps---and each constraint contains two steps, so the time taken to solve $\wspi{1}(\ne)$ is $O(c(k-1)^k)$ and the time taken to solve \mbox{$\wspi{1}(\ne,=)$} is $O(c(k-1)^k + kn)$.

Now suppose we are given an instance of \wsp$(=,\ne)$.
Then we can determine its satisfiability by considering the satisfiability of multiple instances of $\wspi{1}(=,\ne)$, each instance containing $c$ constraints.
We now determine the number of instances of \mbox{$\wspi{1}(=,\ne)$} that need to be considered in the worst case.

For a Type 2 constraint $(\rho,s,S')$, we may assume that $s \not\in S'$: for $(=,s,S')$, if $s \in S'$, then the constraint is satisfied by every plan and the constraint is redundant; for $(\ne,s,S')$, if $s \in S'$, then the constraint is equivalent to $(\ne,s,S' \setminus \set{s})$.
Hence, each Type 2 constraint $(\rho,s,S')$ gives rise to $\card{S'}$ literals in a clause with $\card{S'} < k$.
So we have $c$ clauses, each of which contains no more than $k-1$ literals.

Type 1 constraints are equivalent to clauses with a single literal.
Hence, for an instance of $\wspi{2}(=,\ne)$ there are no more than $(k-1)^c$ simple constraint expressions and so there are no more than $(k-1)^c$ instances of \wsp$_1(=,\ne)$ to check, which can be done in time $O((k-1)^c (c(k-1)^k +  kn + kc)) = O((k-1)^c (c(k-1)^k + kn))$.

Each Type 3 constraint $(\rho,S',S'')$ yields a clause containing fewer than $\card{S'}\cdot\card{S''} \leqslant k^2$ literals (which is greater than the number of clauses that can be obtained from a Type 1 or Type 2 constraint).
Hence, there are no more than $O(k^{2c})$ simple constraint expressions and \wsp$_3(\ne,=)$ can be decided in time $O(k^{2c}(c(k-1)^k +kn + kc)) = (k^{2c}(c(k-1)^k +kn))$.
\end{proof}

\begin{Cor}\label{cor:wsp2-and-wsp3-are-fpt}
\wsp$_2(=,\ne)$ and \wsp$_3(=,\ne)$ are FPT.
\end{Cor}

\subsection{Kernelization of WSP}

Formally, a \emph{parameterized problem} $P$ can be represented as a relation $P\subseteq \Sigma^* \times \mathbb{N}$ over a finite alphabet $\Sigma$.
The second component is call the {\em parameter} of the problem.
In particular, {\sc WSP} is a parameterized problem with parameter $k$, the number of steps.
We denote the size of a problem instance $({\cal I},k)$ by $\card{{\cal I}} + k$.

\begin{Def}
Given a parameterized problem $P$, a \emph{kernelization of $P$} is an algorithm that maps an instance $({\cal I},k)$ to an instance $({\cal I}',k')$ in time polynomial in $|{\cal I}|+k$ such that
  \begin{inparaenum}[(i)]
    \item $({\cal I},k)\in P$ if and only if $({\cal I}',k')\in P$, and
    \item \mbox{$k'+|{\cal I}'|\leqslant g(k)$} for some function  $g$;
  \end{inparaenum}
$({\cal I}',k')$ is the \emph{kernel} and $g$ is the {\em size} of the kernel.
If $g(k)=k^{O(1)}$, then we say $({\cal I}',k')$ is a {\em polynomial-size} kernel.
\end{Def}

A kernelization provides a form of preprocessing aimed at compressing the given instance of the problem.
Polynomial-size kernels are particularly useful in practice as they often allow us to reduce the size of the input of the problem under consideration to an equivalent problem with an input of significantly smaller size.

Crampton {\em et al.} recently established that \mbox{$\wspi{1}(=,\ne)$} has a polynomial-size kernel~\cite[\S 6]{CrGuYe12a}.
In the case of \mbox{$\wspi{1}(=,\ne)$}, we can reduce the problem to one containing at most $k$ users~\cite[Theorem 6.5]{CrGuYe12a}.
Crampton {\em el al.} also showed that $\wspi{2}(=,\ne)$ (and hence $\wspi{3}(=,\ne)$) does not have a polynomial-size kernel, so there is no efficient preprocessing step for such instances of \wsp.
However, our results in this paper show we can reduce an instance of $\wspi{}(=,\ne)$ to at most $k^{2c}$ instances of $\wspi{1}(\ne)$ and then solve each instance by first computing a (polynomial-size) kernel.
The proof of Corollary~\ref{cor:wsp2-and-wsp3-are-fpt} asserts that $c \leqslant 4^k$, although we would expect $c$ to be linear or quadratic in the number of steps in practice.
This approach is similar to those that use so-called \emph{Turing kernels} (see~\cite{ScKoMoNi12}, for example).

\subsection{Negative Constraint Expressions}\label{sec:negative-constraint-expressions}

We could extend the syntax for constraint expressions to include negation.
In other words, if $\gamma$ is a constraint expression, then $\neg \gamma$ is a constraint expression.
A plan $\pi$ satisfies $\neg\gamma$ if and only if $\pi$ violates $\gamma$.
A plan $\pi$ satisfies the constraint $\neg(=,S_1,S_2)$, for example, if and only for all $s_i \in S_i$, $(\pi(s_1),\pi(s_2)) \not\in \Delta$; that is, if and only if for all $s_i \in S_i$, $\pi(s_1) \ne \pi(s_2)$.%
\footnote{This constraint is similar to the separation of duty constraints described by Basin {\em et al.}~\cite{BaBuKa12} and the universal constraints described by Wang and Li~\cite{WangLi10}.
Of course, we can represent this constraint as the set of Type 1 constraints $\set{(\ne,s_1,s_2) : s_1 \in S_1,s_2 \in S_2}$.}
Thus, we can encode any instance of \mbox{$\wspi{}(=,\ne)$} using only constraints of the form $(=,s_1,s_2)$ if we allow the use of negation.
Note, however, this means that the method for solving $\wspi{}(=,\ne)$ described in Section~\ref{sec:solving-wsp3} no longer works, because we may have negative literals in our conjunctive normal form expressions.

However, we can determine the satisfiability of the constraint expression using any SAT solver.
A satisfying assignment returned by the SAT solver provides a ``template'' for a valid plan: if the variable $(=,s_1,s_2)$ is set to true, then our plan must assign the same user to $s_1$ and $s_2$.
This induces a partition of the set of steps into blocks, each of which must be executed by a different user.
Hence, each satisfying assignment of the constraint expression gives rise to an instance of $\wspi{1}(\ne)$ in which each ``step'' is a block of steps in the original problem instance.
We can solve this instance in time $O(\binom{k}{2}(b-1)^b)$, where $b \leqslant k$ is the number of blocks, since there are at most $\binom{k}{2}$ constraints of the form $(=,s_1,s_2)$.
If $b$ is small relative to $k$, then this may prove to be a very efficient way of solving the original instance of $\wspi{}(=,\ne)$.
However, we may need to consider $2^{\binom{k}{2}}$ satisfying assignments.
In future work we hope to explore whether the additional expressive power of negative constraint expressions allows us to encode business rules of practical relevance.
Further experimental work, investigating which strategies for solving \wsp\ work best in practice, is required.

\section{Conditional Workflows}\label{sec:conditional-wsp}

\newcommand{\sinks}[1]{\overline{#1}}
\newcommand{\sources}[1]{\underline{#1}}
\newcommand{\opseries}{\mathbin{;}}
\newcommand{\opparallel}{\mathbin{\parallel}}
\newcommand{\opxor}{\oplus}
\newcommand{\concat}{\rightarrow}
\newcommand{\merge}{\leftrightarrows}

\newcommand{\wfsrl}{\mathsf{serial}}
\newcommand{\wfprl}{\mathsf{parallel}}
\newcommand{\wfxor}{\mathsf{xor}}

\newcommand{\firststep}{\alpha}
\newcommand{\laststep}{\omega}
\newcommand{\midstep}{\epsilon}

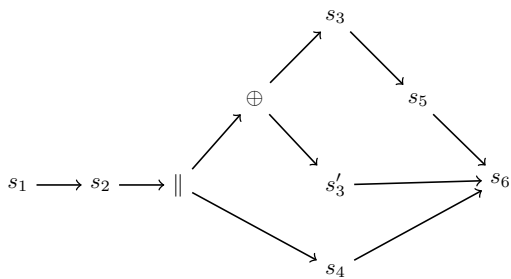
\begin{figure}[h]\centering
\begin{tikzpicture}[->,.=stealth',node distance=8mm and 7mm,semithick,auto,scale=.85,transform shape]
  \node  (t1)                      {$s_1$};
  \node  (t2) [right=of t1]        {$s_2$};
  \node (parallel) [right=of t2] {$\opparallel$};
  \node (xor) [above right=of parallel] {$\opxor$};
  \node  (t3) [above right=of xor]  {$s_3$};
  \node (t3dash) [below right=of xor] {$s'_3$};
  \node  (t4) [below=of t3dash]  {$s_4$};
  \node  (t5) [above right=of t3dash]  {$s_5$};
  \node  (t6) [below right=of t5]  {$s_6$};
  \path (t1) edge (t2)
   (t2) edge (parallel)
   (parallel) edge (xor)
   (parallel) edge (t4)
   (xor) edge (t3)
   (xor) edge (t3dash)
   (t3) edge (t5)
   (t3dash) edge (t6)
   (t4) edge (t6)
   (t5) edge (t6);
\end{tikzpicture}
\caption{A workflow specification with conditional step execution}\label{fig:ext-wkflw-example}
\end{figure}

In some situations, we may wish to have conditional branching in a workflow specification, sometimes known as \emph{OR-forks}~\cite{AaHoKiBa03} or \emph{exclusive gateways}~\cite{WhMi08}.
In our workflow system for purchase order processing, for example, we may require that only orders with a value exceeding some threshold amount need to be signed for twice.
Informally, we can represent this extended specification by the diagram shown in Figure~\ref{fig:ext-wkflw-example}, where $s'_3$ represents a step for signing a goods received note on low-valued items.
The nodes containing $\opparallel$ and $\opxor$ are ``orchestration'' steps (or ``gateways'') at which no processing is performed: $\opxor$ indicates that exactly one of the two branches is executed, while $\opparallel$ denotes that both branches must be executed.

\subsection{Workflow Composition}

We now introduce a simple language for defining workflows.
This language enables us to extend the definition of WSP to workflows containing OR-forks, but not to arbitrary workflow patterns.

We assume every workflow specification includes a start step and a finish step, which we will denote by $\firststep$ and $\laststep$, respectively, with subscripts where appropriate.
These steps are \emph{orchestration} steps: no processing is performed by these steps and no constraints are applied to their execution; they are used by the workflow management system solely to manage the initiation and completion of workflow instances.
Given two workflow specifications $W_1 = (S_1,E_1)$ and $W_2 = (S_2,E_2)$, we may construct new workflow specifications using serial, parallel and xor composition, denoted by $W_1 \opseries W_2$, $W_1 \opparallel W_2$ and $W_1 \oplus W_2$, respectively.
We assume throughout that $S_1 \cap S_2 = \emptyset$.
(If this were not the case with $s \in S_1 \cap S_2$, we could simply introduce subscripts or new labels to distinguish the two copies of $s$.)

For serial composition, all the steps in $W_1$ must be completed before the steps in $W_2$.
Hence, the graph of $W_1 \opseries W_2$ is formed by taking the union of $S_1$ and $S_2$, the union of $E_1$ and $E_2$, and the addition of a single edge between $\laststep_1$ and $\firststep_2$.

For parallel composition, the execution of the steps in $W_1$ and $W_2$ may be interleaved.
Hence, the graph of $W_1 \parallel W_2$ is formed by taking the union of $S_1$ and $S_2$, the union of $E_1$ and $E_2$, the addition of new start and finish steps $\firststep_{\rm par}$ and $\laststep_{\rm par}$, and the addition of edges from $\firststep_{\rm par}$ to $\firststep_1$ and $\firststep_2$ and from $\laststep_1$ and $\laststep_2$ to $\laststep_{\rm par}$.
This form of composition is sometimes known as an \emph{AND-fork}~\cite{AaHoKiBa03} or a \emph{parallel gateway}~\cite{WhMi08}.%
\footnote{The workflows that arise from serial and parallel composition have a lot in common with series-parallel graphs; see~\cite{BaGu08}, for example, for further details.}

In both serial and parallel composition, all steps in $W_1$ and $W_2$ are executed.
In xor composition, either the steps in $W_1$ are executed or the steps in $W_2$, but not both.
In other words, xor composition represents non-deterministic choice in a workflow specification.
The graph of $W_1 \oplus W_2$ is formed by taking the union of $S_1$ and $S_2$, the union of $E_1$ and $E_2$, the addition of new start and finish steps $\firststep_{\rm xor}$ and $\laststep_{\rm xor}$, and the addition of edges from $\firststep_{\rm xor}$ to $\firststep_1$ and $\firststep_2$ and from $\laststep_1$ and $\laststep_2$ to $\laststep_{\rm xor}$.

Henceforth, we will assume that $\laststep_1$ followed by $\firststep_2$ will be merged to form a single (orchestration) node $\midstep$.
Similarly, we will assume that
  \begin{inparaenum}[(i)]
    \item $\firststep_{\rm par}$ followed by $\firststep_1$ and $\firststep_2$ in serial composition will be merged to form a single node $\firststep_{\rm par}$;
    \item $\laststep_{\rm par}$ followed by $\laststep_1$ and $\laststep_2$ will be merged to form a single node $\laststep_{\rm par}$;
    \item $\firststep_{\rm xor}$ followed by $\firststep_1$ and $\firststep_2$ will be merged to form a single node $\firststep_{\rm xor}$;
    \item $\laststep_{\rm xor}$ followed by $\laststep_1$ and $\laststep_2$ will be merged to form a single node $\laststep_{\rm xor}$.
  \end{inparaenum}

\begin{figure}[h]\centering
\subfigure[Serial]{
\begin{tikzpicture}[->,.=stealth',node distance=8mm and 7mm,semithick,auto,state/.style={draw,circle,inner sep=3pt},scale=.85,transform shape]
  \node  (start1)        {$\firststep_1$};
  \node (w1) [right=of start1]  {$W_1$};
  \node (mid1) [right=of w1] {$\midstep$};
  \node (w2) [right=of mid1] {$W_2$};
  \node (end2) [right=of w2] {$\laststep_2$};
  \path (start1) edge (w1)
   (w1) edge (mid1)
   (mid1) edge (w2)
   (w2) edge (end2);
\end{tikzpicture}}
\hfill
\subfigure[Parallel]{
\begin{tikzpicture}[->,.=stealth',node distance=8mm and 7mm,semithick,auto,state/.style={draw,circle,inner sep=3pt},scale=.85,transform shape]
  \node (startpar) {$\firststep_{\rm par}$};
  \node (w1) [above right=of startpar]  {$W_1$};
  \node (w2) [below right=of startpar] {$W_2$};
  \node (endpar) [below right=of w1] {$\laststep_{\rm par}$};
  \path (startpar) edge (w1)
   (startpar) edge (w2)
   (w1) edge (endpar)
   (w2) edge (endpar);
\end{tikzpicture}}
\caption{Workflow composition}\label{fig:workflow-composition}
\end{figure}
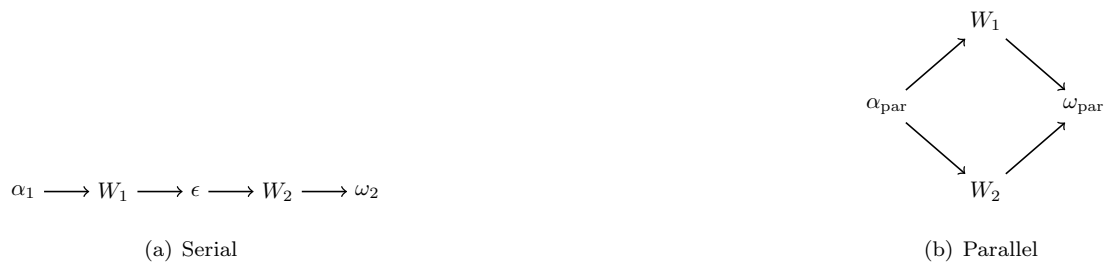

Serial and parallel composition are illustrated in Figure~\ref{fig:workflow-composition}.
The structure of xor composition is identical to that for parallel composition so it is not shown.

\subsection{Execution Sets}

When we have conditional branching in a workflow, there exists more than one set of steps that could comprise a complete workflow instance.
Formally, an \emph{execution set} is defined recursively:
  \begin{itemize}
    \item for a workflow specification comprising a single step $s$, there is a single execution set $\set{s}$;
    \item if $W_1$ and $W_2$ are workflow specifications and $S_1$ and $S_2$ are execution sets for $W_1$ and $W_2$, respectively, then
  \begin{itemize}
    \item $S_1 \cup S_2$ is an execution set for $W_1 \opseries W_2$,
    \item $S_1 \cup S_2$ is an execution set for $W_1 \opparallel W_2$,
    \item $S_1$ and $S_2$ are execution sets for $W_1 \opxor W_2$.
  \end{itemize}
  \end{itemize}
In our running example, both $\set{s_1,s_2,s_3,s_4,s_5,s_6}$ and $\set{s_1,s_2,s'_3,s_4,s_6}$ represent possible execution sets, with the second set representing a workflow instance in which the value of goods ordered is lower than the threshold requiring the GRN to be countersigned.\footnote{The concept of an execution set is related to, but simpler than, the concept of an \emph{execution history}~\cite{BaBuKa12}: for any execution set $\set{s_1,\dots,s_m}$, an execution history is a set $\set{(s_1,u_1),\dots,(s_m,u_m)}$ for some users $u_1,\dots,u_m$. An execution history also has some similarity to our concept of a plan.}

\begin{oldstuff}
Note that we could also define inclusive-or composition (the operation corresponding to an \emph{inclusive gateway} orchestration step in BPMN~\cite{WhMi08}), which we will denote by $+$.
In this case, one or more of the sub-workflows must be executed.
Hence, if $S_1$ and $S_2$ are execution sets for $W_1$ and $W_2$, respectively, then $S_1$, $S_2$ and $S_1 \cup S_2$ are executions sets for $W_1 + W_2$.
Hence, every possible subset of $S$ can be an execution set.
We omit consideration of inclusive-or for the remainder of this paper.
\end{oldstuff}

\subsection{Workflow Formulas and Trees}

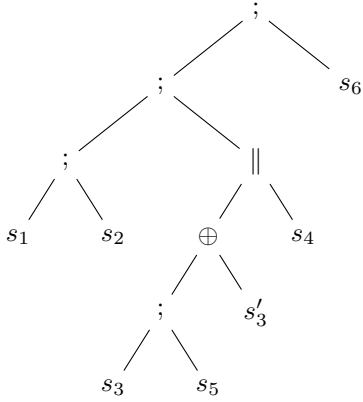
\begin{figure}[h]\centering
\begin{tikzpicture}[grow=down, sloped,auto]
\node {$\opseries$}
    child {
        node {$\opseries$}
            child {
                node {$\opseries$}
                  child {
                    node (s1) {$s_1$}
                  }
                  child {
                    node (s2) {$s_2$}
                  }
            }
            child {
                node {$\opparallel$}
                  child {
                    node {$\opxor$}
                      child {
                        node {$\opseries$}
                          child {
                            node (s3) {$s_3$}
                          }
                          child {
                            node (s5) {$s_5$}
                          }
                      }
                      child {
                        node (s3new) {$s'_3$}
                      }
                  }
                  child {
                    node (s4) {$s_4$}
                  }
            }
    }
    child {
      node (s6) {$s_6$}
    };
\end{tikzpicture}
\caption{A workflow tree} \label{fig:wrkflw-tree}
\end{figure}

Clearly, each workflow step represents a workflow specification, in fact the simplest possible specification.
Hence, we may represent the example workflow specification in Figure~\ref{fig:ext-wkflw-example} as the \emph{workflow formula}
  \[
    (s_1 \opseries s_2) \opseries (((s_3 \opseries s_5) \opxor s'_3) \opparallel s_4) \opseries s_6.
  \]
Thus, we may also represent the workflow specification as a \emph{workflow tree}, as illustrated in Figure~\ref{fig:wrkflw-tree}.

\tikzstyle{level 1}=[level distance=1cm, sibling distance=2.5cm]
\tikzstyle{level 2}=[level distance=1cm, sibling distance=2.5cm]
\tikzstyle{level 3}=[level distance=1cm, sibling distance=1.25cm]

The number of different possible execution sets is determined by the structure of the workflow formula.
Specifically, let $\sharp(W)$ denote the number of possible execution sets for workflow $W$.
For a workflow $W$ comprising a single step, we have $\sharp(W) = 1$.
In general, we have
  \begin{align*}
    \sharp(W_1 \opseries W_2) &= \sharp(W_1 \opparallel W_2) = \sharp(W_1) \cdot \sharp(W_2) \\
    \sharp(W_1 \opxor W_2) &= \sharp(W_1) + \sharp(W_2),
  \end{align*}
where $\cdot$ denotes multiplication.

Using a post-order traversal of the workflow tree, we can compute the number of possible execution sets: we assign the value $1$ to each leaf node; we compute the number of possible execution sets for each non-leaf node using the values assigned to its children and the appropriate formula for the operation associated with the node.
The root node in the tree depicted in Figure~\ref{fig:wrkflw-tree} is assigned the value $2$, for example.

We write $\flat(W)$ to denote the maximum number of steps in any possible execution set for a workflow specification $W$.
Then
  \begin{align*}
    \flat(W_1 \opseries W_2) &= \flat(W_1 \opparallel W_2) = \flat(W_1) + \flat(W_2) \\
    \flat(W_1 \opxor W_2) &= \max\set{\flat(W_1),\flat(W_2)}.
  \end{align*}
Clearly, we can compute $\flat(W)$ from the workflow tree associated with $W$ using a similar algorithm to the one described above for calculating $\sharp(W)$.

\subsection{Constraints in Conditional Workflows}\label{sec:constraints-conditional-workflows}

Let $W_1$ and $W_2$ be two workflow specifications with constraints $C_1$ and $C_2$, respectively.
When we form $W_1 \opseries W_2$ or $W_1 \opparallel W_2$, we include all constraints in $C_1$ and $C_2$.
In addition, we may create new constraints, governing the execution of some steps in $S_1$ and some steps in $S_2$.
However, we prohibit the addition of constraints in which all the steps are contained in either $S_1$ or $S_2$ (the assumption being that they would have been created earlier, if required).
In other words, any constraint that is added when we form $W_1 \opseries W_2$ (or $W_1 \opparallel W_2$) has the form $(\rho,S',S'')$, where $S' \cup S'' \not \subseteq S_1$ and $S' \cup S'' \not \subseteq S_2$.

In contrast, since xor composition requires that we either perform the steps in $S_1$ or those in $S_2$, any constraint that includes steps from both $S_1$ and $S_2$ serves no purpose.
Hence, we assume that we add no constraints when we form $W_1 \opxor W_2$.

\subsection{Derived Deterministic Workflows}

We say a workflow specification is \emph{deterministic} if it has a single execution set (and \emph{non-deterministic} otherwise).
Each possible execution set in a non-deterministic workflow specification gives rise to a different, deterministic workflow specification.
In particular, given a workflow specification $W = (S,E)$ with execution sets $\set{S_1,\dots,S_m}$, we define $W_i = (S_i,E_i)$, where \[ E_i \stackrel{\rm def}{=} (S_i \times S_i) \cap E. \]
Then $W_i$ is a (derived) deterministic workflow specification.

For a constrained workflow specification $W = ((S,E),A,C)$ with possible execution sets $\set{S_1,\dots,S_m}$, we define $W_i = (S_i,E_i,A_i,C_i)$, where \[ A_i \stackrel{\rm def}{=} (S_i \times U) \cap A,\] and, for each $\gamma = (\rho,S_1,S_2) \in C$ such that $S_1 \cap S_i \ne \emptyset$ and $S_2 \cap S_i \ne \emptyset$, \[ \gamma_i \stackrel{\rm def}{=} (\rho,S_1 \cap S_i,S_2 \cap S_i) \in C_i.\]
Each $W_i$ is a deterministic, constrained workflow specification.
Notice that when we form $\gamma_i$, $S_1 \cap S_i \ne \emptyset$ and $S_2 \cap S_i \ne \emptyset$: this follows by a simple induction on the structure of the workflow formula and the assumptions we make about the addition of constraints when we compose workflows (as described in Section~\ref{sec:constraints-conditional-workflows}).

Hence, we may model any non-deterministic workflow specification as a collection of deterministic workflow specifications.
We may define the notion of \emph{weakly satisfiable} and \emph{strongly satisfiable} for a non-deterministic specification: the former holds if there exists a derived, deterministic workflow specification that is satisfiable; the latter holds if all derived, deterministic workflow specifications are satisfiable.
In practice, it is likely that a workflow specification should be strongly satisfiable (otherwise there exist execution paths that can never complete).

\begin{Pro}\label{pro:wsp-conditional-fpt}
Let $W$ be an instance of $\wspi{1}(=,\ne)$.
Then we can determine whether $W$ is weakly or strongly satisfiable in time \mbox{$O(\sharp(W)(\flat(W)-1)^{\flat(W)})$}.
\end{Pro}

Note that we can extend this result to $\wspi{3}(=,\ne)$ as described in the proof of Theorem~\ref{thm:wsp2-elementary} (that is, using the reduction to multiple instances of $\wspi{1}(=,\ne)$, where the number of instances is $O(\flat(W)^2)$).
The above result asserts that the complexity of checking whether a workflow is strongly satisfiable is determined by $\flat(W)$ and $\sharp(W)$.
Crude upper bounds for these parameters are $k$ and $2^k$, both functions of $k$ only.
Thus, determining whether a conditional workflow is strongly satisfiable is FPT.

Of course, these bounds can be improved: the upper bound for $\flat(W)$ is only attained if no xor composition is used, in which case $\sharp(W) = 1$; conversely, introducing xor composition may reduce the maximum length, and using only xor composition reduces the number of derived specifications to $k$.
The question is: What deployment of $k-1$ composition operators for $k$ steps yields the worst-case complexity?
We have the following result.

\begin{Thm}\label{thm:execution-sets}
Given $k$ workflow steps, a workflow has no more than:
  \begin{itemize}
    \item $3^{k'}$ execution sets if $k= 3k'$;
    \item $4 \cdot 3^{k' - 1}$ execution sets if $k = 3k'+1$; and
    \item $2 \cdot 3^{k'}$ execution sets if $k = 3k'+2$.
  \end{itemize}
\end{Thm}

\begin{Rem}
The proof of the above result (see appendix) is constructive, in the sense that it tells us how to maximize the number of execution sets for a fixed set of $k$ steps.
Given $k$ steps, we obtain a workflow with the greatest possible number of execution sets by taking the serial (or parallel) composition of sub-workflows $\oplus_2$ and $\oplus_3$, where $\oplus_i$ denotes the xor composition of $i$ steps.
More specifically, if $k = 3a$, we take the serial composition of $a$ copies of $\oplus_3$; if $k = 3a+1$, we take the serial composition of $a-1$ copies of $\oplus_3$ and two copies of $\oplus_2$; and if $k = 3a+2$, we take the serial composition of $a$ copies of $\oplus_3$ and one copy of $\oplus_2$.
We may conclude that $\flat(W)$ for such a workflow is no greater than $\ceil{k/3}$.
\end{Rem}

\begin{Rem}
Note that using xor composition reduces $\flat(W)$.
And note that the exponential term in the complexity of solving $\wspi{1}(=,\ne)$ is determined by the number of steps in the workflow, for which an upper bound is $\flat(W)$ in the case of non-deterministic workflow specifications.
For a fixed $k$, it follows from Theorem~\ref{thm:execution-sets} that the worst-case complexity for $\wspi{1}(=,\ne)$ occurs for a workflow specification with a single execution set (of $k$ steps).
\end{Rem}

\section{Further Applications}\label{sec:other-applications}

In this section, we study two problems from the literature and establish that they are fixed-parameter tractable.
In both cases, we represent the problem as a workflow satisfiability problem using constraint expressions.

\subsection{Ordered \wsp}\label{sec:owsp}

We note that the version of \wsp\ considered so far in this paper makes no use of the order relation on the set of steps.
This is a simplification introduced by Wang and Li~\cite{WangLi10}.
In fact, the definition of workflow constraints by Crampton~\cite{cram:sacmat05} prohibited constraints of the form $(\rho,s,s')$ for $s > s'$.
Moreover, a plan was required to specify an execution order for the steps in the workflow (in addition to the assignment of steps to users).
This, in turn, means that Crampton's definition of constraint satisfaction (and hence of the workflow satisfiability problem) is more complex.
More formally, we have the following definitions.

\begin{Def}
Let $\mathcal{W} = ((S,E),U,A,C)$ be a workflow comprising $k$ steps.
A tuple $(s_1,\dots,s_k)$ is an \emph{execution schedule} for $\cal W$ if $\set{s_1,\dots,s_k} = S$ and, for all $1 \leqslant i < j \leqslant k$, $s_i \not\geqslant s_j$.\footnote{In other words, an execution schedule is a \emph{linear extension} or \emph{topological sort} of $(S,\leqslant)$.}
We say $s_i$ \emph{precedes} $s_j$ in an execution schedule if $i < j$.
\end{Def}

For the workflow depicted in Figure~\ref{fig:example-workflow}, $(s_2,s_1,\dots)$ is not an execution schedule, for example, but $(s_1,s_2,s_3,s_5,s_4,s_6)$ and $(s_1,s_2,s_3,s_4,s_5,s_6)$ are.

\begin{Def}\label{def:owsp-constraint-satisfaction}
The (Type $1$) constraint $(\rho,s,s')$ is \emph{satisfied} by execution schedule $\sigma$ and plan $\pi$ if one of the following holds:
  \begin{inparaenum}[(i)]
    \item $s$ precedes $s'$ in $\sigma$ and $(\pi(s),\pi(s')) \in \rho$;
    \item $s'$ precedes $s$ in $\sigma$.
  \end{inparaenum}
\end{Def}

The intuition here is that a constraint $(\rho,s,s')$ is well-formed only if $s$ could precede $s'$ in the execution of some instance of the workflow (that is, either $s < s'$ or $s \parallel s'$).
Moreover, if $s$ does occur before $s'$, then the execution of $s'$ is constrained by $\rho$ and the identity of the user that performed $s$.
A modified version of \wsp, based on the above definitions, is defined in the following way.

\begin{center}
\fbox{%
      \begin{tabulary}{.95\columnwidth}{@{}r<{~}@{}L@{}}
        \multicolumn{2}{@{}l}{\sc Ordered WSP (\owsp)}\\
        \emph{Input:} & A constrained workflow authorization schema $((S,E),U,A,C)$. \\
        \emph{Output:} & {\sc True} if there exists an execution schedule $\sigma$ and a plan $\pi$ that satisfy all constraints in $C$, and {\sc False} otherwise.
       \end{tabulary}%
      }
\end{center}

Note that it may not be possible to find a valid plan $\pi$ for a particular execution schedule $\sigma$.
Conversely, there may be a plan $\pi$ for which there exist schedules $\sigma$ and $\sigma'$ such that $(\sigma,\pi)$ satisfies all constraints but $(\sigma',\pi)$ does not.
Consider, for example, a plan $\pi$ that is valid for our purchase order workflow such that $\pi(s_3) = \pi(s_4)$.
If we add the constraint $(\ne,s_3,s_4)$, then $\pi$ is valid for any execution schedule in which $s_4$ precedes $s_3$ and invalid otherwise.
%

The above example also shows there exist workflows for which a plan $\pi$ is not a solution to {\sc WSP}, but for which $(\sigma,\pi)$ is a solution to \owsp\ for certain choices of $\sigma$.
Crampton introduced the notion of a \emph{well-formed workflow}, which has the following property: for all $s_i \parallel s_j$, $(\rho,s_i,s_j) \in C$ if and only if $(\tilde{\rho},s_j,s_i) \in C$, where $\tilde{\rho}$ is defined to be $\set{(u,u') \in U \times U : (u',u) \in \rho}$.
To ensure that the workflow in the above example is well-formed, we would add the constraint $(\ne,s_4,s_3)$ to $C$.
It is easy to see that \owsp\ for well-formed workflows and WSP are essentially equivalent, since a valid plan for one execution schedule will be a valid plan for any execution schedule~\cite[Lemma 9]{cram:sacmat05}.

Nevertheless, there will be business processes that cannot be represented using a well-formed workflow schema.
In the purchase order example illustrated in Figure~\ref{fig:example-workflow}, for example, it would be quite reasonable to impose constraints on $s_3$ and $s_4$ that would mean the resulting workflow schema was not well-formed.
Suppose, for example, that $\sim$ is an equivalence relation on $U$, where $u \sim u'$ if and only if $u$ and $u'$ belong to the same department.
Then the constraints $(\nsim,s_3,s_4)$ and $(\ne,s_4,s_3)$ require that if $s_3$ (the sign GRN step) is performed before $s_4$ (the create payment step), then the user that performs $s_4$ must be in a different department from the user that performs $s_3$; whereas if the steps are performed in the reverse order, we only require the users to be different (since the more commercially sensitive step has been performed first in this case).

Note that \owsp\ is only defined for Type 1 constraints (see Definition~\ref{def:owsp-constraint-satisfaction}).
Wang and Li showed that \wsp\ is W[1]-hard~\cite{DowneyFellows99} for arbitrary constraint relations (even if only Type 1 constraints) are used~\cite{WangLi10}.
Moreover, any instance of \wsp\ defines an instance of \owsp.
Thus, \owsp\ is W[1]-hard.
However, there is a strong connection between \wsp\ and \owsp.

\newcommand{\owspi}[1]{\text{OWSP}_{#1}}

\begin{Pro}\label{pro:owsp-is-fpt}
$\owspi{1}(\rho_1,\dots,\rho_t)$ is FPT if $\wspi{1}(\rho_1,\dots,\rho_t)$ is.
\end{Pro}

A stronger notion of satisfiability for OWSP would require that there exists a plan for every execution schedule (as for conditional workflows).
In this case, we simply require that every one of the $O(k!)$ derived instances of WSP is satisfiable.
The worst-case complexity of determining ``weak'' and ``strong'' satisfiability  for OWSP is, therefore, the same.
Note that an instance of WSP is satisfiable if the corresponding instance of OWSP is strongly satisfiable.


\subsection{Identifying Constraint Violation}

Consider the following problem: {\em Given a constrained workflow specification $((S,E),U,A,C)$, does there exist a plan $\alpha$ such that $(s,\alpha(s)) \in A$ for all $s \in S$ and at least one constraint $C$ that is not satisfied?}
This question is of interest because if we know that no such plan exists, then we do not need a reference monitor: any allocation of (authorized) users to steps will satisfy all the constraints.
This question has been studied by Armando and colleagues~\cite{ArGiPo09,ArRa11} and solutions have been computed using model checkers.
We answer this question by examining the satisfiability of the ``negation'' of the problem, rewritten using the language of constraint expressions.

\begin{Thm}\label{thm:constraint-violation}
Determining whether there exists a plan that violates a workflow specification $((S,E),U,A,C)$, where all constraints have the form $(=,S_1,S_2)$ or $(\ne,S_1,S_2)$, is FPT.
\end{Thm}

The approach described above can also be used to ``prune'' a workflow specification.
Given a workflow specification $((S,E),U,A,C)$, we can identify, with the same (worst-case) time complexity, all constraints in $C$ that can be violated.
This enables us to remove any constraints that cannot be violated, leaving a workflow specification $((S,E,U,A,C')$, with $C' \subseteq C$.
In Section~\ref{sec:applications-of-wsp}, we identified situations in which we may be required to solve WSP for a workflow specification multiple times.
Thus, reducing the set of constraints will reduce the complexity of subsequent attempts to determine the satisfiability of the workflow specification.


\section{Concluding Remarks} \label{sec:conclusion}

In this paper, we have explored the use of constraint expressions as a means of translating different versions of the workflow satisfiability problem into one or more instances of $\wspi{1}(\ne)$.
Constraint expressions provide a uniform way of representing the workflow satisfiability problem and related problems, such as WSP for conditional workflows (Section~\ref{sec:conditional-wsp}), ordered WSP and the identification of constraints that can be violated (Section~\ref{sec:other-applications}).
This, in turn, enables us to establish the complexity of solving these problems.
We also believe our characterization of workflow composition, the representation of workflows as trees, and execution sets may be useful modeling tools for future research on authorization in workflow systems.

\subsection{Related Work}

Work on computing plans for workflows that must simultaneously satisfy authorization policies and constraints goes back to the seminal paper of Bertino {\em et al.}~\cite{BeFeAt99}.
This work considered linear workflows and noted the existence of an exponential algorithm for computing valid plans.
Crampton extended the model for workflows to partially ordered sets (equivalently, directed acyclic graphs) and to directed acyclic graphs with loops~\cite{cram:sacmat05}.
Wang and Li further extended this model to include Type $2$ constraints~\cite{WangLi10}.

Wang and Li first investigated the computational complexity and, significantly, the existence of fixed-parameter tractable algorithms for the workflow satisfiability problem~\cite{WangLi10}.
One or their main results~\cite[Theorem 9]{WangLi10} is very similar to the result we prove for $\wspi{2}(=,\ne)$ (Theorem~\ref{thm:wsp2-elementary}), although our approach is more direct and generalizes to $\wspi{3}(=,\ne)$.
Crampton {\em et al.} introduced a new method for solving the problem~\cite{CrGuYe12}, which yields significantly better complexity bounds for $\wspi{3}(=,\ne)$.
However, their methods only apply for certain kinds of constraints; indeed, it is not clear whether their approach extends to relations other than $\Delta$, $\Delta^c$ and constraints using equivalence relations defined on the user set.

The use of constraint expressions to represent and reason about the complexity of the workflow satisfiability problem appears, therefore, to have some significant advantages, one specific advantage being its versatility, over existing approaches.
Khan and Fong introduced the notion of a constraint expression to reason about the problem of workflow feasibility~\cite{KhFo12}, which asks: Given a set of constraints and restrictions on admissible authorization policies, does there exist an authorization policy from which we can construct a valid plan?
Their work was undertaken in the context of the relationship-based access control model~\cite{Fong11}, in which the ``shape'' of authorization policies is restricted, and does not explore fully the possibility of using constraints expressions to solve the ``classical'' workflow satisfiability problem.

It is widely accepted that it is useful to have conditional branching in workflow specifications~\cite{AaHoKiBa03,WhMi08}.
However, there is very little prior work on the workflow satisfiability problem, or its complexity, for conditional workflows.
Khan's master's thesis includes work on existential satisfiability (what we have called weak satisfiability) and universal (strong) satisfiability~\cite[Chapter 8]{khan:sati12} but does not consider fixed parameter tractability.

\subsection{Future Work}

There are a number of opportunities for future work.
Crampton {\em et al.} studied the workflow satisfiability problem in the presence of constraints specified using an equivalence relation $\sim$ defined on $U$~\cite{CrGuYe12}.
The relation $\Delta$ may be viewed as an equivalence relation, in which each equivalence class is a single user.
We would like to investigate whether our methods can be extended to solve $\wspi{}(=,\ne,\sim,\nsim)$, where $\sim$ is not equal to $\Delta$.
This is a non-trivial problem as we cannot use our trick of considering only those steps for which there are fewer than $k$ authorized users.
A second problem we would like to consider is the \emph{optimal workflow-aware authorization administration problem}, which determines whether it is possible to modify the authorization relation, subject to some bound on the ``cost'' of the changes, when the workflow is unsatisfiable~\cite{BaBuKa12}.
Finally, we would like
\begin{inparaenum}[(a)]
  \item to remove the restriction that $(S,E)$ is an ayclic graph, so that we can model sub-workflows that can be repeated, and
  \item to include \emph{inclusive gateways}~\cite{WhMi08}, allowing for one or more sub-workflows to be executed.
\end{inparaenum}
Both of these extensions can be readily modeled using execution sets (or multisets).
If, for example, $S_1$ and $S_2$ are execution sets for $W_1$ and $W_2$, respectively, then $S_1$, $S_2$ and $S_1 \cup S_2$ are execution sets for $W_1 + W_2$, where $+$ indicates inclusive-or composition.

\bibliography{refs}
\bibliographystyle{abbrv}

\vfill

\appendix\section{Proofs}

\begin{proof}[Proof of Corollary~\ref{cor:wsp2-and-wsp3-are-fpt}]
For $\wspi{2}(=,\ne)$, in the worst case, each constraint has the form $(\rho,s,S')$, with $s \not\in S'$.
Hence, the number of Type 2 constraints can be no greater than $k 2^{k-1}$.
It now follows from Theorem~\ref{thm:wsp2-elementary} that $\wspi{2}(=,\ne)$ is FPT.
For $\wspi{3}(=,\ne)$, in the worst case, each constraint has the form $(\rho,S',S'')$.
Thus, noting that $(\rho,S',S'')$ is equivalent for $\wspi{3}(=,\ne)$, the number of Type 3 constraints can be no greater than $2^k \cdot 2^k = 2^{2k}$, from which it follows that $\wspi{3}(=,\ne)$ is FPT.
\end{proof}

\begin{proof}[Proof of Proposition~\ref{pro:wsp-conditional-fpt}]
The result follows by noting that determining strong satisfiability requires us to check whether all $\sharp(W)$ derived instances of $W$ are satisfiable, while determining weak satisfiability requires us to check whether at least one derived instance is satisfiable. The complexity, in the worst case, is the same.
The complexity of checking a single instance is $(k'-1)^{k'}$, where $k'$ is the number of steps in the derived instance.
The result now follows.
\end{proof}

\begin{proof}[Proof of Theorem~\ref{thm:execution-sets}]
First observe that may disregard the $\parallel$ operator in computing an upper bound on $\sharp(W)$.
To see this, note that the parallel operator requires, like the serial operator, that all steps in the sub-workflows are performed.
In particular, an execution set for workflow $W_1 \opparallel W_2$ has the form $S_1 \cup S_2$, where $S_i$ is an execution set for workflow $W_i$.

Recall $\oplus_i$ represents the xor composition of $i$ steps and $\sharp(\oplus_i \opseries \oplus_j) = \sharp(\oplus_j \opseries \oplus_i) = ij$.

We proceed by induction on $k$.
For $k=2$, we may construct
  \[
    \text{$\oplus_1 \opseries \oplus_1\quad\text{and}\quad\oplus_2$},
  \]
thus the result holds for $k=2$.
For $k=3$, we may construct three different workflows:
  \[
    \oplus_1 \opseries \oplus_1 \opseries \oplus_1,\quad\oplus_1 \opseries \oplus_2,\quad\text{and}\quad\oplus_3,
  \]
thus the result holds for $k=3$.
Finally, for $k=4$, we may construct
  \[
    \oplus_1 \opseries \oplus_1 \opseries \oplus_1 \opseries \oplus_1,\quad\oplus_1 \opseries \oplus_1 \opseries \oplus_2,\quad
    \oplus_1 \opseries \oplus_3,\quad \oplus_2 \opseries \oplus_2\quad\text{and}\quad\oplus_4,
  \]
thus the result holds for $k=4$.

Now consider $k > 4$ steps and suppose the result holds for all workflows constructed from $k-1$ or fewer steps.
Then for any split of $k$ into workflows $W_1$ and $W_2$ comprising $k_1$ and $k_2$ steps, respectively, such that $k_1 + k_2 = k$, we may form $W_1 \opseries W_2$ or $W_1 \opxor W_2$.
Clearly, for $k > 4$, $\sharp(W_1 \opseries W_2) > \sharp(W_1 \opxor W_2)$.
Moreover, $\sharp(W_1 \opseries W_2) = \sharp(W_1) \cdot \sharp(W_2)$.

First consider the case $k = 3a$ and let $k_i = 3a_i + b_i$, $i = 1,2$, with $b_i \in \set{0,1,2}$.
We assume (without loss of generality) that $b_1 \leqslant b_2$.
If $b_1 = b_2$, then $k_1$ and $k_2$ are divisible by $3$ and $\sharp(W_i) \leqslant 3^{a_i}$ by the inductive hypothesis, whence \[ \sharp(W) = \sharp(W_1) \cdot \sharp(W_2) \leqslant 3^{a_1} \cdot 3^{a_2} = 3^{a}. \]
If $b_1 = 1$ and $b_2 = 2$, then we have $a_1 + a_2 = a-1$ and
  \[
    \sharp(W) = \sharp(W_1) \cdot \sharp(W_2) \leqslant 4 \cdot 3^{a_1 - 1} \cdot 2 \cdot 3^{a_2} = 8 \cdot 3^{a-2} < 3^{a}
  \]
and the result holds.

Now consider the case $k = 3a+1$.
If $b_1 = 0$, then $b_2 = 1$ and $a_1 + a_2 = a$.
Hence, by the inductive hypothesis, we have
\[
  \sharp(W) \leqslant 3^{a_1} \cdot 4 \cdot 3^{a_2-1} = 4 \cdot 3^{a-1},
\]
as required.
If $b_1 = b_2$, then we have $b_i = 2$ and $a_1 + a_2 = a - 1$.
Hence, by the inductive hypothesis, we have
\[
  \sharp(W) \leqslant 2 \cdot 3^{a_1} \cdot 2 \cdot 3^{a_2} = 4 \cdot 3^{a-1},
\]
as required.

Finally, consider the case $k = 3a+2$.
If $b_1 = 0$, then $b_2 = 2$ and $a_1 + a_2 = a$.
Hence, by the inductive hypothesis, we have
\[
  \sharp(W) \leqslant 3^{a_1} \cdot 2 \cdot 3^{a_2} = 2 \cdot 3^a,
\]
as required.
If $b_1 = b_2$, then $b_i = 1$ and $a_1 + a_2 = a$.
Hence, we have
\[
  \sharp(W) \leqslant 16 \cdot 3^{a_1 + a_2 - 2} = \frac{16}{9} \cdot 3^a < 2 \cdot 3^a.
\]
as required.
\end{proof}

\begin{proof}[Proof of Proposition~\ref{pro:owsp-is-fpt}]
An instance of $\owspi{1}$ contains a set of constraints $C$ and we may assume that $C$ contains at least two constraints of the form $(\rho_i,s,s')$ and $(\rho_j,s',s)$ with $\rho_i \ne \widetilde{\rho_j}$.
(If no such constraints exist then $\owspi{1}(\rho_1,\dots,\rho_t)$ is identical to an instance of $\wspi{1}(\rho_1,\dots,\rho_t)$.)
Observe that the number of linear extensions of $(S,\leqslant)$ (and hence possible execution schedules) is determined only by $k$.
Specifically, the number of linear extensions is no greater than $k!$.
Note also that in any execution of the workflow, either $s$ precedes $s'$ or vice versa.
Hence each linear extension allows us to discard either $(\rho_i,s,s')$ or $(\rho_j,s',s)$ (since exactly one of them will be irrelevant to the schedule defined by the linear extension), thus defining an instance of \wsp\ that contains fewer constraints than the original problem.
In other words, we may consider our instance of $\owspi{1}$ to be the disjunction of $k!$ instances of $\wspi{1}$.
If each instance of $\wspi{1}$ is FPT, we can solve each of these instances, thus solving the original instance of $\owspi{1}$.
\end{proof}

\begin{proof}[Proof of Theorem~\ref{thm:constraint-violation}]
A Type 1 constraint $(\rho,s,s')$ is satisfied by a plan $\alpha$ if $(\alpha(s),\alpha(s')) \in \rho$ and is not satisfied (``violated'') otherwise.
In other words, $(\rho,s,s')$ is violated by $\alpha$ if $(\alpha(s),\alpha(s')) \not\in \rho$.
Equivalently, a constraint $(\rho,s,s')$ is violated iff $(\overline{\rho},s,s')$ is satisfied, where
  \[
    \overline{\rho} \stackrel{\rm def}{=} \set{(u,u') \in U \times U : (u,u') \not\in \rho}.
  \]

A Type 2 constraint $(\rho,s,S')$, $S' \subseteq S$ is violated if $(\rho,s,s')$ is violated for all $s' \in S'$.
In other words, $(\rho,s,S')$ is violated iff the constraint expression
  \[
    \bigwedge_{s' \in S'} (\overline{\rho},s,s')
  \]
is satisfied.
Similarly, a Type 3 constraint $(\rho,S',S'')$ is violated iff the constraint expression
  \[
    \bigwedge_{s' \in S',s'' \in S''} (\overline{\rho},s',s'')
  \]
is satisfied.
Finally a set of constraints $\set{c_1,\dots,c_t}$ is violated if at least one $c_i$ is violated.
In other words, we can determine whether there exists a plan that violates a set of constraints by determining if there exists a plan $\alpha$ that satisfies a constraint expression in disjunctive normal form, where each clause is a conjunction of Type 1 constraints.
We make the following observations.
\begin{itemize}
  \item There are no more than $c$ disjuncts, where $c$ is the number of constraints in the original workflow specification.
  \item A Type 2 constraint, when rewritten in the above way, gives rise to a conjunction of no more than $k-1$ Type 1 constraints, while a Type 3 constraint gives rise to no more than $k^2$ Type 1 constraints.
  \item There can be no more than $k 2^k$ Type 2 constraints in a workflow specification and no more than $4^k$ Type 3 constraints.
  \item $\overline{\Delta}$ is $\Delta^c$ and $\overline{\Delta^c}$ is $\Delta$.
  \item By Theorem~\ref{thm:wsp2-elementary}, the time taken to solve $\wspi{1}(\Delta,\Delta^c)$ (that is, $\wspi{1}(=,\ne)$) is $O(c(k-1)^k + kn)$, where $c$ is the number of constraints.
\end{itemize}

Therefore, there exists an FPT algorithm to determine whether there exists a plan $\pi$ in which each user is authorized and a constraint that $\pi$ does not satisfy, since we need only find a single disjunct that is true, and each disjunct represents a workflow specification containing only Type 1 constraints.
The time taken to solve this new problem is $O(k 2^{k-1}((k-1)^{k+1} + kn))$ for Type 2 constraints and $O(4^k (k(k-1)^{k+1} + kn))$ for Type 3 constraints.
\end{proof}

\end{document}